Tree Inference:  Response Time in a Binary Multinomial Processing Tree,
Representation and Uniqueness of Parameters

Richard Schweickert, Xiaofang Zheng

Department of Psychological Sciences
Purdue University
West Lafayette, IN

Author Note

Acknowledgements:  This work is supported in part by a scholarship from China Scholarship Council (CSC) to Zheng.

Correspondence concerning this article should be sent to Richard Schweickert, Department of Psychological Sciences, Purdue University, 703 Third St., West Lafayette, IN 47907, schweick@purdue.edu.

Abstract

A Multinomial Processing Tree (MPT) is a directed tree with a probability associated with each arc. Here we consider an additional parameter associated with each arc, a measure such as the time required to select the arc. MPTs are often used as models of tasks. Each vertex represents a process and an arc descending from a vertex represents selection of an outcome of the process. A source vertex represents processing that begins when a stimulus is presented and a terminal vertex represents making a response. Responses are partitioned into classes. An experimental factor selectively influences a vertex if changing the level of the factor changes parameter values on arcs descending from that vertex and on no others. Earlier work shows that if each of two experimental factors selectively influences a different vertex in an arbitrary MPT it is equivalent for the factors to one of two relatively simple MPTs. Which of the two applies depends on whether the two selectively influenced vertices are ordered by the factors or not. A special case, the Standard Binary Tree for Ordered Processes, arises if the vertices are so rdered and the factor selectively influencing the first vertex changes parameter values on only two arcs descending from that vertex. Here we derive necessary and sufficient conditions for the probability and measure associated with a particular response class to be accounted for by this special case. Parameter values are not unique and we give admissible transformations for transforming one set of parameter values to another. When an experiment with two factors is conducted, the number of observations and parameters to be estimated depend on the number of levels of each factor; we provide degrees of freedom.

Tree Inference:  Response Time in a Binary Multinomial Processing Tree,
Representation and Uniqueness of Parameters

A Multinomial Processing Tree (MPT) is a tree with parameters on its arcs, see Figure 1. The tree consists of a finite set of vertices and a set of arcs, each arc being an ordered pair of vertices, such that there is no more than one directed path from one vertex to another. Associated with each arc is a probability parameter and possibly other parameters.  The sum of the probabilities on arcs descending from a vertex is 1.  A source vertex is a vertex with no incoming arc, that is, a vertex that is not the second vertex of any arc.  A terminal vertex is a vertex with no outgoing arc, that is, a vertex that is not the first vertex of any arc.

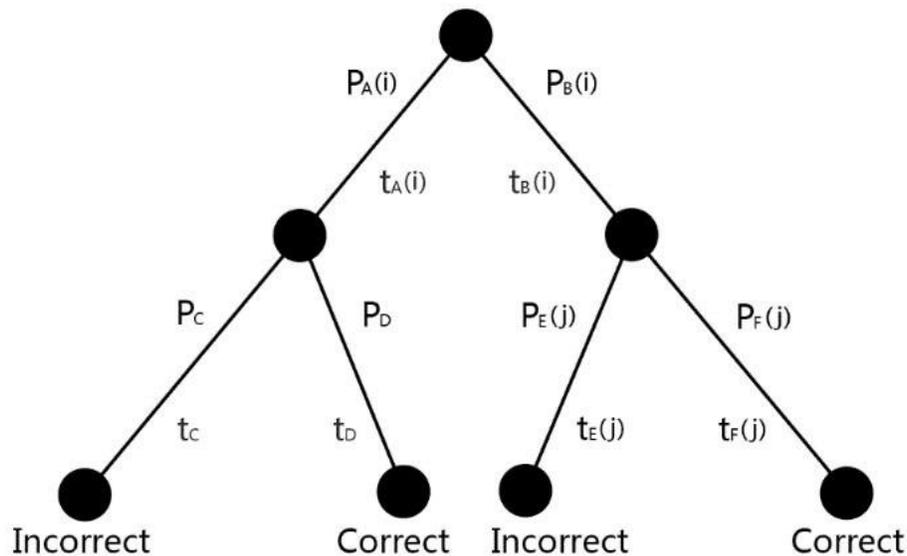

Figure 1.  A Multinomial Processing Tree.  It is the Standard Binary Tree for Ordered Processes.

Multinomial Processing Trees are widely used to model cognitive processing in perception, memory, decision making, and movement tasks.  For reviews, see Batchelder and Riefer (1999), Erdfelder, Auer, Hilbig, Aβfalg, Moshagen and  Nadarevic (2009), and Hütter and Klauer (2016).  In an MPT model of a task, each vertex represents a process, for example, an attempt to retrieve an item from memory.  The process has possible outcomes, such as successful or unsuccessful retrieval.  Each possible outcome is represented by a vertex with an arc directed to it from the vertex representing the process.  The probability associated with the arc is the probability the corresponding outcome occurs after the process starts.  When the task starts, with, say, presentation of a stimulus, processing begins at a source vertex.  One of the outcomes of the source vertex occurs, with the probability associated with the corresponding arc.  The second vertex of the arc is reached and the process represented by that vertex starts.  Steps continue in this fashion until a terminal vertex is reached, at which a response is made.  Responses are partitioned into classes, such as correct and incorrect.  One class of responses is made at each terminal vertex, although several terminal vertices may be associated with the same response class.

An MPT may have several sources. Usually, these correspond to different conditions. For example, one source may be for stimuli that are verbs, another for nouns, and in a given condition only one source is relevant and the others can be ignored. We assume the MPTs under consideration have only one source.

The probability of a directed path from one vertex to another is the product of the probabilities associated with the arcs on the path. On any particular trial of the task, a directed path is followed from the source to a terminal vertex. The probability of a response of a particular class is the sum of the probabilities of paths from the source to the terminal vertices of that class.

When a response is made, an observation might be made of the time required to make the response, or of some other quantity. To account for these, response times and other measures are sometimes incorporated in MPTs (e. g., Heck & Erdfelder, 2016; Hu, 2001, Klauer & Kellen, 2018; Link, 1982; Schweickert & Zheng, 2018; Wollschläger & Diederich, 2012). Here we assume that each arc has associated with it, in addition to a probability, another parameter. We often consider the additional parameter to be the time required for the outcome corresponding to the arc to occur, but the parameter could have another interpretation, such as a cost, so we call it a measure. We assume the measure for a directed path from one vertex to another is the sum of the measures associated with the arcs on the path.

Selective influence

An experimental factor, such as the brightness of a stimulus, selectively influences a vertex if changing a level of the factor changes parameter values on some arcs descending from the vertex and on no other arcs. We sometimes say the factor selectively influences the process that the vertex represents. For a review of selective influence in MPTs see Schweickert, Fisher and Sung (2012).

Suppose an experiment is carried out with two factors, $\Phi$ and $\Psi$. A level of Factor $\Phi$ is denoted $i$, with $i = 1, \ldots, I$ and a level of Factor $\Psi$ is denoted $j$, with $j = 1, \ldots, J$. We are interested in the situation in which each factor selectively influences a different vertex in an MPT. For example, in Figure 1, the source vertex is selectively influenced by Factor $\Phi$ and a vertex following the source vertex is selectively influenced by Factor $\Psi$. Parameters on arcs descending from the source vertex are indexed by the level of Factor $\Phi$. When Factor $\Phi$ is at level $i$, the probability the vertex on the left is the outcome of the processing at the source vertex is $P_A(i)$ and $t_A(i)$ is the measure associated with this outcome (e.g., the time required for this outcome to occur). Other notation is similar.

For the MPT in Figure 1, when Factor $\Phi$ is at level $i$ and Factor $\Psi$ is at level $j$ the probability $p(i,j)$ of a correct response is
$$p(i,j) = P_A(i)P_D + P_B(i)P_F(j).$$
The measure $t(i,j)$ associated with a correct response, given a correct response is made, is a mixture of measures of each of the two paths to a terminal vertex for a correct response. That is,
$$t(i,j) = [P_A(i)P_D/p(i,j)][t_A(i) + t_D] + [P_B(i)P_F(j)/p(i,j)][t_B(i) + t_F(j)].$$
More conveniently,

$$p(i,j)t(i,j) = P_A(i)P_D[t_A(i) + t_D] + P_B(i)P_F(j)[t_B(i) + t_F(j)].$$

By saying every level of a factor selectively influencing a vertex is effective we mean each factor has at least two levels and there are no two levels of a factor such that when the levels of all other factors are fixed, in every response class the probability of a response is the same for both levels and the additional measure is the same for both levels. (It is possible that for two levels of a factor in some response class the probability of a response is the same for both levels, but the measure changes, or vice versa.)

Vertex arrangements

Consider an MPT in which each of two factors selectively influences a different vertex. There are only two ways the two vertices can be arranged in the MPT. Suppose there is a directed path from the source vertex to a terminal vertex, and on this path there is an arc whose parameter values depend on the level of one of the factors, and also an arc whose parameter values depend on the level of the other factor. Then we say the vertices are ordered by the factors, or for short, ordered. If there is no such path, we say the vertices are unordered by the factors, or for short, unordered. (Note that a path directed from one selectively influenced vertex to the other will not suffice for the vertices to be ordered by the factors, if no arc on the path has a parameter whose value depends on a level of one of the factors.) In the MPT in Figure 1 the selectively influenced vertices are ordered by the factors.

Two MPTs are equivalent for Factors $\Phi$ and $\Psi$, with respective levels $i = 1, \ldots, I$ and $j = 1, \ldots, J$, if the MPTs lead to the same values of $p(i,j)$ and $t(i,j)$ for every $i$ and $j$. Earlier work shows that if each of two factors selectively influences a different vertex in an arbitrary MPT with two response classes, that MPT is equivalent to one of two relatively simple MPTs (Schweickert & Zheng, 2019a, 2019b). If in the arbitrary MPT the selectively influenced vertices are ordered by the factors, the MPT is equivalent to the Standard Tree for Ordered Processes in Figure 2. Otherwise, the arbitrary MPT is equivalent to the Standard Tree for Unordered Processes in Figure 3.

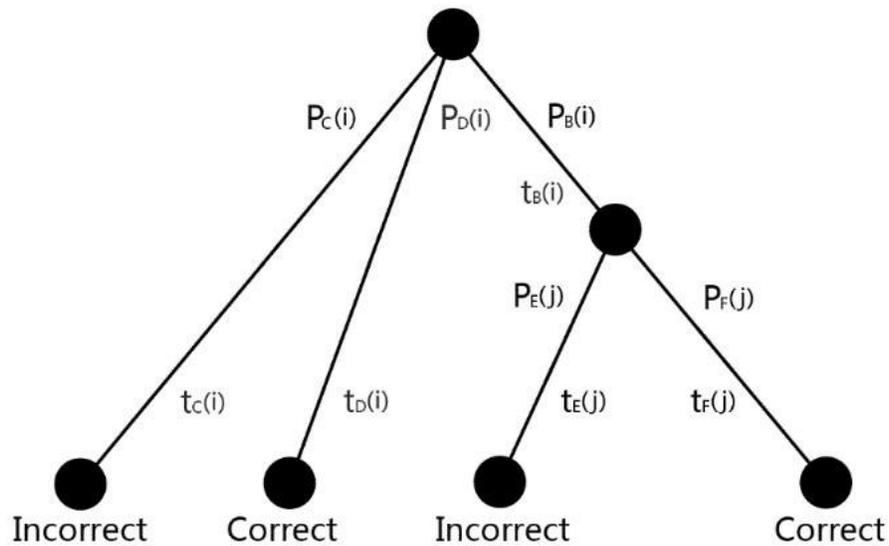

Figure 2. The Standard Tree for Ordered Processes.

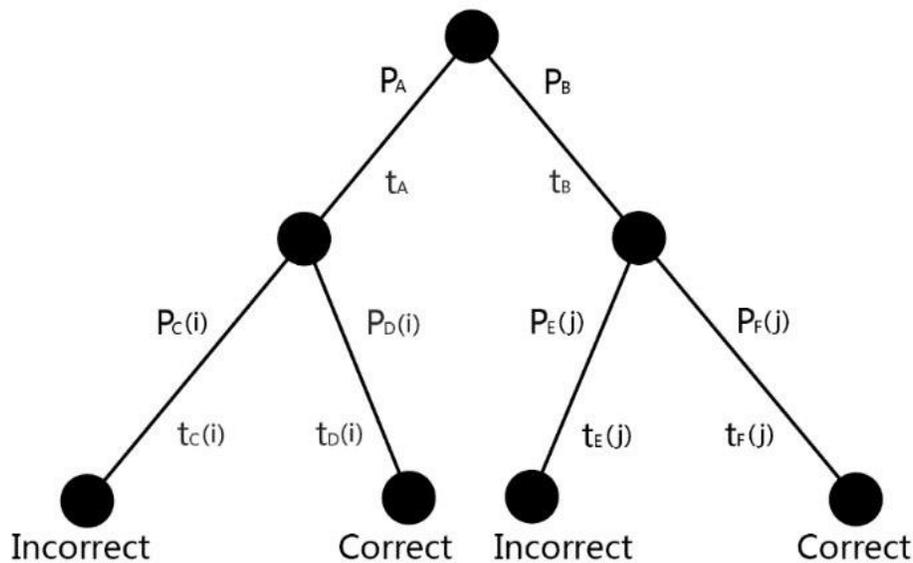

Figure 3. The Standard Tree for Unordered Processes.

Three arcs descend from the source of the Standard Tree for Ordered Processes in Figure 2. In many MPT models the tree is binary, that is, exactly two arcs descend from every nonterminal vertex. It is useful to consider the special case of binary trees. One reason is that any MPT representing response probability can be reparameterized as a binary MPT (Hu & Batchelder, 1994; Knapp & Batchelder, 2004). Further, a way of representing binary MPTs as strings in a context-free language has been developed by Purdy and Batchelder (2009), facilitating analysis of model structure.

In what follows, we assume there are exactly two response classes. For binary trees there is nothing new to consider for selectively influenced vertices that are unordered by the factors, because the Standard Tree for Unordered Processes is itself a binary tree. For vertices ordered by the factors, consider the binary tree in Figure 1, with exactly two arcs descending from each nonterminal vertex. It turns out that if in an arbitrary MPT with two response classes, two factors selectively influence vertices $v_1$ and $v_2$ ordered by the factors, with $v_1$ preceding $v_2$, the arbitrary MPT is equivalent to the MPT in Figure 3, provided a simple condition is met. The condition is that the factor selectively influencing vertex $v_1$ changes parameters on only two descending arcs. This was shown for response probabilities by Schweickert and Chen (2008), and, when an additional measure is included, by Schweickert and Zheng (2019b).

The MPT in Figure 1 is called the *Standard Binary Tree for Ordered Processes*. The theorem to follow gives necessary and sufficient conditions for it to account for response probability and an additional measure. By saying every level of a factor selectively influencing a vertex is effective we mean each factor has at least two levels and there are no two levels of a factor such that when the level of all the other factors are fixed, in every response class the probability of a response is the same for both levels and the additional measure is the same for both levels. (It is possible for two levels of a factor that in some response class the probability of a response is the same for both levels, but the measure changes, or vice versa.) We assume there are two response classes, which we label as correct and incorrect. Suppose Factor $\Phi$ is at level $i$ and Factor $\Psi$ is at level $j$. The probability of a correct response is denoted $p(i,j)$. The probability of an incorrect response is $1 - p(i,j)$. The measure associated with a correct response is denoted $t(i,j)$ and that associated with an incorrect response is denoted $t_w(i,j)$.

**Definition**. Probability matrix $\mathbf{P} = \big(p(i,j)\big)$, correct-response-measure matrix $\mathbf{T} = \big(t(i,j)\big)$, and incorrect-response-measure matrix $\mathbf{T_w} = \big(t_w(i,j)\big)$ are produced by two factors selectively influencing two vertices ordered by the factors in the Standard Binary Tree for Ordered Processes, with the vertex selectively influenced by Factor $\Phi$ preceding the vertex selectively influenced by Factor $\Psi$, if the following are true. Both factors are effective and for all $i$ and $j$ there are probability parameters $p_A(i)$, $p_B(i)$, $p_C$, $p_D$, $p_E(j)$, $p_F(j)$ such that

$$0 \le p_A(i), p_B(i), p_C, p_D, p_E(j), p_F(j) \le 1$$
$$p_A(i) + p_B(i) = 1$$
$$p_C + p_D = 1$$
$$p_E(j) + p_F(j) = 1$$

and

$$p(i,j) = p_A(i)p_D + p_B(i)p_F(j).$$

Further, there are measure parameters $t_A(i)$, $t_B(i)$, $t_D$, $t_F(j)$ such that

$$p(i,j)t_1(i,j) = p_A(i)p_D[t_A(i) + t_D] + p_B(i)p_F(j)[t_B(i) + t_F(j)].$$

And there are further measure parameters $t_C$ and $t_E(j)$ such that

$$[1 - p(i,j)]t_w(i,j) = p_A(i)[1 - p_D][t_A(i) + t_C] + p_B(i)p_E(j)[t_B(i) + t_E(j)].$$

The following theorem gives necessary and sufficient conditions, testable with data, for response probabilities and response measures to be produced by factors selectively influencing vertices in the Standard Binary Tree for Ordered Processes. We note that response probabilities $p(i,j)$ are required to be strictly between 0 and 1 for all levels $i$ and $j$ of the factors, to avoid dividing by 0 at certain places in the proof. The proof uses a result from Schweickert and Zheng (2019b, Theorem 5), which is provided in the appendix here.

**Theorem 1.** *Suppose for all i and j, $0 < p(i,j) < 1$. Probability matrix $\mathbf{P} = \big(p(i,j)\big)$, correct-response measure matrix $\mathbf{T} = \big(t(i,j)\big)$, and incorrect-response measure matrix $\mathbf{T_w} = \big(t_w(i,j)\big)$ are produced by Factor $\Phi$ and Factor $\Psi$ selectively influencing two different vertices ordered by the factors in the Standard Binary Tree for Ordered Processes, with the vertex selectively influenced by Factor $\Phi$ preceding the vertex selectively influenced by Factor $\Psi$, if and only if*
*there is a level n of Factor $\Psi$ and for every level i of Factor $\Phi$ there are numbers $r_i \geq 0$ and $s_i$, such that the following three conditions are true.*

    1. *There is a constant k, $0 \leq k \leq 1$, such that for every $i \neq h$ and $j \neq n$*
$$p(i,j)p(h,n) - p(i,n)p(h,j) = k[p(i,j) - p(h,j) - p(i,n) + p(h,n)].$$

    2. *For every j,*
$$p(i,j) - k = r_i[p(h,j) - k].$$

    3. *Let $max\{r_i\} = r_h$. For every j,*
$$r_h r_i s_i [\, p(h, j) - p(h,n)]$$
$$= r_h[\, p(i, j)t(i, j) - p(i,n)t(i,n)] - r_i[\, p(h, j)t(h, j) - p(h,n)t(h,n)].$$
$$= - r_h\{[1 - p(i,j)]t_w(i,j) - [1 - p(i,n)]t_w(i,n)]\}$$
$$+ r_i\{[1 - p(h,j)]t_w(h,j) - [1 - p(h,n)]t_w(h,n)]\}. \tag{1}$$

**Proof.** I. Suppose probability matrix $\mathbf{P} = (p(i,j))$, $\mathbf{0} < \mathbf{P} < \mathbf{1}$, and measure matrix $\mathbf{T} = (t(i,j))$ are produced by Factor $\Phi$ and Factor $\Psi$ selectively influencing two different vertices ordered by the factors in the Standard Binary Tree for Ordered Processes, with the vertex selectively influenced by Factor $\Phi$ preceding the vertex selectively influenced by Factor $\Psi$.

    Then for any i and j,
$$p(i, j) = p_A(i)p_D + p_B(i)p_F(j) \tag{2}$$
and
$$p(i, j)t(i, j) = p_A(i)p_D[t_A(i) + t_D] + p_B(i)p_F(j)[t_B(i) + t_F(j)],$$
with $p_A(i) = 1 - p_B(i)$.

    Let h be a value of i such that max $\{p_B(i)\} = p_B(h)$. Let n be a value of j such that min $\{p(h,j)t(h,j)\} = p(h,n)t(h,n)$.

    With a little algebra it follows from Equation (2) that for every $i \neq h$ and $j \neq n$
$$p(i,j)p(h,n) - p(i,n)p(h,j) = p_D[p(i,j) - p(h,j) - p(i,n) + p(h,j)].$$
Hence, Condition 1 is true, with $k = p_D$.

Note that if $p_B(h) = 0$, then for every $i$, $p_B(i) = 0$ so Factor $\Phi$ is ineffective. Hence $p_B(h) \neq 0$. For every $i$ let $r_i = p_B(i) / p_B(h)$. Clearly, $0 \leq r_i$.

From Equation (2), for every $i$ and for every $j$,

$$p(i, j) - k = k + p_B(i)[p_F(j) - k] - k$$
$$= [p_B(i) / p_B(h)]p_B(h)[p_F(j) - k]$$
$$= r_i[p(h, j) - k].$$

Hence, Condition 2 is true.

We turn to Condition 3.

If $r_i = 0$, let $s_i = 0$.

For $i$ such that $r_i \neq 0$, let $s_i = t_B(i) - t_B(h)$. Clearly, $s_i$ does not depend on $j$. We now consider the right hand side of Equation (1) and show that it equals $r_h r_i s_i[p(h, j) - p(h, n)]$.

Suppose $r_i = 0$. Then $s_i = 0$, so $r_h r_i s_i[p(h, j) - p(h, n)] = 0$.

Because $r_i = p_B(i) / p_B(h) = 0$, it follows that $p_B(i) = 0$ and $p_A(i) = 1$. The right hand side of Equation (1) is also 0, that is,

$$r_h[p(i, j)t(i, j) - p(i, n)t(i, n)]$$
$$= r_h[p_A(i)p_D - p_A(i)p_D] = 0.$$

Suppose $r_i \neq 0$.

Select $j$ such that $p(h, j) \neq p(h, n)$. Because neither $r_h$, $r_i$ nor $p(h,j) - p(h,n)$ is 0, we can divide expressions by $r_h r_i[p(h,j) - p(h,n)]$.

On division by $r_h r_i[p(h,j) - p(h,n)]$, the left hand side of Equation (1) is $s_i$.

On division by $r_h r_i[p(h,j) - p(h,n)]$, the right hand side of Equation (1) is

$$\frac{[p(i, j)t(i, j) - p(i, n)t(i, n)]/r_i - [p(h, j)t(h, j) - p(h, n)t(h, n)]/r_h}{p(h, j) - p(h, n)}. \quad (3)$$

The first term in the numerator of Equation (3) is

$$[p(i, j)t(i, j) - p(i, n)t(i, n)]/r_i = \{p_B(i)p_F(j)[t_B(i) + t_F(j)] - p_B(i)p_F(n)[t_B(i) + t_F(n)]\}/r_i$$

$$= p_B(h)\{p_F(j)[t_B(i) + t_F(j)] - p_F(n)[t_B(i) + t_F(n)]\}.$$

Similarly, the second term in the numerator of Equation (3) is

$[p(h,j)t(h,j) - p(h,n)t(h,n)]/r_h = p_B(h)\{p_F(j)[t_B(h) + t_F(j)] - p_F(n)[t_B(h) + t_F(n)]\}.$

The numerator of Equation (3) becomes

$p_B(h)\{p_F(j)[t_B(i) - t_B(h)] - p_F(n)[t_B(i) - t_B(h)]\}$

$= p_B(h)[p_F(j) - p_F(n)][t_B(i) - t_B(h)].$

The denominator of Equation (3) is

$$p(h,j) - p(h,n) = p_B(h)[p_F(j) - p_F(n)].$$

Then the ratio in Equation (1) becomes

$$t_B(i) - t_B(h) = s_i.$$

Hence, Condition 3 is true.

II. Suppose Conditions 1, 2 and 3 are true. We begin by showing that matrices **P** and **T** are produced by the Standard Tree for Ordered Processes, and then we show they are also produced by the Standard Binary Tree for Ordered Processes.

Schweickert and Chen (2008) showed that from Conditions 1 and 2, parameters $p_A(i)$, $p_B(i)$, $p_C$, $p_D$, $p_E(j)$, and $p_F(j)$ exist, such that

$$0 \leq p_A(i), p_B(i), p_C, p_D, p_E(j), p_F(j) \leq 1$$
$$p_A(i) + p_B(i) = 1$$
$$p_C + p_D = 1$$
$$p_E(j) + p_F(j) = 1$$

and

$$p(i,j) = p_A(i)p_D + p_B(i)p_F(j), \qquad (4)$$

with $p_D = k$.

We now show that **P** and **T** are produced by Factor $\Phi$ and Factor $\Psi$ selectively influencing two vertices in the Standard Tree for Ordered Processes. Necessary and sufficient conditions are in Schweickert and Zheng (2019b) Theorem 5 (see Appendix).

Renumber the levels of Factor $\Psi$ so $p_F(1) \leq \ldots \leq p_F(j) \leq \ldots \leq p_F(J)$. Then by Equation (4), for every $i$, $p(i,1) \leq \ldots \leq p(i,j) \leq \ldots \leq p(i,J)$. Hence, Condition 1 of Theorem 5 is satisfied.

By Condition 2 of the theorem to be proved, for every level $i$ of Factor $\Phi$ there exists a number $r_i \geq 0$ such that for every level $j$,

$$p(i,j) - k = r_i[p(h,j) - k],$$

where $h$ is a level such that $\max\{r_i\} = r_h$. Let $n$ be a level of Factor $\Psi$ such that $\min\{p(h,j)t(h,j)\} = p(h,n)t(h,n)$. Then for every $i$ and $j$,

$$p(i,j) - p(i,n) = p(i,j) - k - [p(i,n) - k]$$
$$= r_i[p(h,j) - k] - r_i[p(h,n) - k]$$
$$= r_i[p(h,j) - p(h,n)].$$

Hence Condition 2 of Theorem 5 is satisfied (with $i* = h$ and $j* = n$).

Condition 3 of Theorem 5 is satisfied because it is the same as Condition 3 of the theorem to be proved, and we have assumed Condition 3 to be true.

Because Conditions 1, 2 and 3 of Theorem 5 are satisfied, there exist probability parameters (denoted $\varphi$) and measure parameters (denoted $\mu$) of the Standard Tree for Ordered Processes such that the following equations are true for every level $i$ of Factor $\Phi$ and every level $j$ of Factor $\Psi$. The arc denoted $D$ in Theorem 5 (for the Standard Tree for Ordered Processes) is denoted $\Delta$ here to distinguish it from arc $D$ in the Standard Binary Tree for Ordered Processes.

$$p(i,j) = \varphi_\Delta(i) + \varphi_B(i)\varphi_F(j). \tag{5}$$
$$p(i,j)t(i,j) = \varphi_\Delta(i)\mu_\Delta(i) + \varphi_B(i)\varphi_F(j)[\mu_B(i) + \mu_F(j)]. \tag{6}$$

From the proof of Theorem 5, parameters can be assigned so $\mu_F(n) = 0$.

Equation (4) has the form of an equation accounting for $p(i,j)$ with the Standard Tree for Ordered Processes. Then Equations (4) and (5) provide two sets of probability parameters that account for $p(i,j)$ with the Standard Tree for Ordered Processes. Then by Theorem 6 of Schweickert and Zheng (2019b), with scaling parameters $c$ and $d$ we can transform probability parameters in Equation (5) to those in Equation (4), as follows.

For every level $i$, $p_B(i) = \varphi*_B(i) = c\varphi_B(i)$.

For every level $i$, $p_A(i)p_D = \varphi*_\Delta(i) = \varphi_\Delta(i) - cd\varphi_B(i)$.

For every level $j$, $p_F(j) = \varphi*_F(j) = \varphi_F(j)/c + d$.

Now by Theorem 6, the following transformations of the measure parameters in Equation (6) are admissible transformations, with scaling parameters $c$ and $d$, and with $j' = n$ and $e = 0$.

For every level $i$, let $\mu*_B(i) = \mu_B(i)$.

Recall that $\mu_F(n) = 0$.

Let $\mu*_F(n) = \mu_F(n) = 0$.

For every level $i$, let

$$\mu_\Delta^*(i) = \frac{-\varphi_\Delta(i)\mu_\Delta(i) + cd\varphi_B(i)\mu_B(i)}{cd\varphi_B(i) - \varphi_\Delta(i)}$$

For every level $j$, let

$$\mu_F^*(j) = \frac{\varphi_F(j)\mu_F(j)}{cd + \varphi_F(j)}$$

Then by Theorem 6, with the transformed parameters, for every level *i* and every level *j*,

$$p(i,j) = \varphi^*_A(i) + \varphi^*_B(i)\ \varphi^*_F(j)$$
$$= p_A(i)p_D + p_B(i)p_F(j)$$

and

$$p(i,j)t(i,j) = \varphi^*_A(i)\mu^*_A(i) + \varphi^*_B(i)\varphi^*_F(j)[\ \mu^*_B(i) + \mu^*_F(j)]$$
$$= p_A(i)p_D[\mu^*_A(i) + 0] + p_B(i)p_F(j)[\ \mu^*_B(i) + \mu^*_F(j)].$$

Then **P** and **T** are produced by the Standard Binary Tree for Ordered Processes, with $t_A(i) = \mu^*_A(i)$, $t_D = 0$, $t_B(i) = \mu^*_B(i)$ and $t_F(j) = \mu^*_F(j)$.

The subtree of the Standard Binary Tree for Ordered Processes that produces correct responses has the same form as the subtree that produced incorrect responses. So, reasoning for **P** and **T_w** is similar to that above.

□

**Uniqueness of parameters.** Suppose the Standard Binary Tree for Ordered Processes accounts for observed response probabilities and response measures with a particular set of parameter values. Those parameter values are not necessarily the only ones that can account for the data.

*Numerical example.* Table 1 gives two different sets of values, old and new, for parameters of the Standard Binary Tree for Ordered Processes that make the same predictions. The parameter values are for a particular level *i* of Factor *Φ* and a particular level *j* of Factor *Ψ*. When Factors *Φ* and *Ψ* have levels *i* and *j*, respectively, the old parameter values predict for the probability of a correct response

$$p(i,j) = p_A(i)p_D + p_B(i)p_F(j) = .50 \text{ x } .40 + .50 \text{ x } .16 = .28.$$

The new parameter values predict the same

$$p(i,j) = p^*_A(i)p^*_D + p^*_B(i)p^*_F(j) = .20 \text{ x } .40 + .80 \text{ x } .25 = .28.$$

Likewise, the old parameter values predict for the product of correct response probability and response measure

$$p(i,j)t(i,j) = p_A(i)p_D[t_A(i) + t_D] + p_B(i)p_F(j)[t_B(i) + t_F(j)] = 2.26.$$

And the new parameter values predict the same

$$p(i,j)t(i,j) = p^*_A\ p^*_D\ [t^*_A(i) + t^*_D] + p^*_B(i)p^*_F(j)[t^*_B(i) + t^*_F(j)] = 2.26.$$

Table 1

*Numerical Example of Transformed Parameters in*

*The Standard Binary Tree for Ordered Processes*

|  | old | new |
|---|---|---|
| $p_A(i)$ | 0.50 | 0.20 |
| $p_B(i)$ | 0.50 | 0.80 |
| $p_D$ | 0.40 | 0.40 |
| $p_F(j)$ | 0.16 | 0.25 |
| $t_A(i)$ | 4.50 | 7.50 |
| $t_D$ | 4.00 | 7.00 |
| $t_B(i)$ | 2.00 | 3.00 |
| $t_F(j)$ | 5.00 | 2.50 |

*Note:* New parameter values were obtained from old ones, including $p_F(j') = .20$ and $t_F(j') = 8$, through the admissible transformations in Table 2, using scaling parameter values $c = 1$, $e = 3$, $f = 1$ and $t*_F(j') = 4$.

Old and new parameter values are related through the admissible transformations in Table 2. There are four scaling parameters, $c, e, f,$ and $t*_F(j')$, where $j'$ is an arbitrary level of Factor $\Psi$, chosen so $p*_F(j') \neq 0$. To insure that new probability parameters are between 0 and 1, the scaling parameters must satisfy the bounds in Table 3, which were derived by Schweickert and Chen (2008).

Table 2

*Admissible Transformations of Parameters:  Standard Binary Tree for Ordered Processes*

---

$p*_B (i) = c p_B (i)$

$p*_D = p_D$

$p*_F (j) = p_F (j)/c + (c-1) p_D /c$

If $p*_B(i) \neq 0$, $t*_B (i) = t_B (i) + f$

If $p*_F(j) \neq 0$,

$$t*_F (j) = \frac{p_F(j) t_F(j) + f \cdot [p_F(j') - p_F(j)] + t*_F (j')[p_F(j') + (c-1)p_D] - p_F(j') t_F(j')}{p_F(j) + (c-1)p_D}$$

If $p*_D \neq 0$, $t*_D = t_D + e$

If $p*_A(i) \neq 0$,

$$t*_A (i) = \frac{p_A(i) t_A(i) + p_B(i)\{[t_B(i) - t_D](1-c) + [f + t*_F (j')][1 - c - \frac{p_F(j')}{p_D}] + \frac{p_F(j') t_F(j')}{p_D} + ce\} - e}{1 - c p_B(i)}$$

If $p*_C \neq 0$, $t*_C = t_C + k$

If $p*_E(j) \neq 0$,

$$t*_E(j) = \frac{p_E(j) t_E(j) + f[p_E(j') - p_E(j)] + [c - p_F(j') - (c-1)p_D] t*_E(j') - p_E(j') t_E(j')}{c - p_F(j) - (c-1)p_D}$$

---

*Note:*  Level $j'$ of Factor $\Psi$ is chosen so $0 \neq p*_F(j') \neq 1$.

Table 3

*Bounds on Scaling Parameters in Admissible Transformations*

*For the Standard Binary Tree for Ordered Processes*

*For probability parameters to be between 0 and 1*

---

$$0 < c \leq 1 / \max\{ p_B(i) \}$$
$$p_D - \min\{ p_F(j) \} \leq p_D \cdot c$$
$$\max\{ p_F(j) \} - p_D \leq (1 - p_D)c$$

---

Suppose response probability and response measure are accounted for by two factors selectively influencing two vertices in the Standard Binary Tree for Ordered Processes. If two sets of parameter values are possible, the following theorem gives the relations between them.

**Theorem 2.** *Suppose probability matrix **P** and measure matrix **T** are produced by Factor Φ and Factor Ψ selectively influencing two different vertices ordered by the factors in the Standard Binary Tree for Ordered Processes, with the vertex selectively influenced by Factor Φ preceding the vertex selectively influenced by Factor Ψ, with probability parameters $p_A(i)$, $p_D$, $p_B(i)$, and $p_F(j)$, and measure parameters $t_A(i), t_B(i), t_D,$ and $t_F(j)$.*

*Then **P** and **T** are produced by Factor Φ and Factor Ψ selectively influencing two different vertices ordered by the factors in the Standard Binary Tree for Ordered Processes, with the vertex selectively influenced by Factor Φ preceding the vertex selectively influenced by Factor Ψ, with probability parameters $p*_A(i)$, $p*_D$, $p*_B(i)$ and $p*_F(j)$, and measure parameters $t*_A(i)$, $t*_D$, $t*_B(i)$ and $t*_F(j)$ if and only if there are constants c, e and f and a level j' of Factor Ψ such that the admissible transformations in Table 2 apply and the bounds in Table 3 are satisfied.*

**Proof.** Suppose probability matrix **P** = $(p(i,j))$ and measure matrix **T** = $(t(i,j))$ are produced by Factor Φ and Factor Ψ selectively influencing two different vertices ordered by the factors in the Standard Binary Tree for Ordered Processes, with the vertex selectively influenced by Factor Φ preceding the vertex selectively influenced by Factor Ψ, with probability parameters $p_A(i)$, $p_D$, $p_B(i)$ and $p_F(j)$, and measure parameters $t_A(i)$, $t_D$, $t_B(i)$ and $t_F(j)$.

I. Suppose **P** and **T** = $(t(i,j))$ are also produced by Factor Φ and Factor Ψ selectively influencing two different vertices ordered by the factors in the Standard Binary Tree for Ordered Processes, with the vertex selectively influenced by Factor Φ preceding the vertex selectively influenced by Factor Ψ, with probability parameters $p*_A(i)$, $p*_D$, $p*_B(i)$ and $p*_F(j)$, and measure parameters $t*_A(i)$, $t*_D$, $t*_B(i)$ and $t*_F(j)$.

The admissible transformations in Table 2, with bounds in Table 3, for the probability parameters $p*_A(i)$, $p*_D$, $p*_B(i)$ and $p*_F(j)$ were shown to apply by Schweickert and Chen (2008).

We turn to admissible transformations for the measure parameters.

Suppose $p*_A(i)$, $p*_D$, $p*_B(i)$, $p*_F(j)$, $t*_A(i), t*_B(i), t*_D$, and $t*_F(j)$ exist with

$$p(i,j)t(i,j) = p*_A(i)p*_D[t*_A(i)+t*_D] + p*_B(i)p*_F(j)[t*_B(i)+t*_F(j)]$$

Also, for any $i$ and $j$,

$$p(i,j)t(i,j) = p_A(i)p_D[t_A(i)+t_D] + p_B(i)p_F(j)[t_B(i)+t_F(j)]$$

Let $j$ and $j'$ be two different values of $j$. Then

$$p(i,j)t(i,j) - p(i,j')t(i,j')$$

$$= p_A(i)p_D[t_A(i)+t_D] + p_B(i)p_F(j)[t_B(i)+t_F(j)] - p_A(i)p_D[t_A(i)+t_D]$$

$$\quad - p_B(i)p_F(j')[t_B(i)+t_F(j')]$$

$$= p_B(i)p_F(j)[t_B(i)+t_F(j)] - p_B(i)p_F(j')[t_B(i)+t_F(j')] \qquad (7)$$

Additionally, for the same $i$ and $j$, due to the assumption of

$$p(i,j)t(i,j) = p*_A(i)p*_D[t*_A(i)+t*_D] + p*_B(i)p*_F(j)[t*_B(i)+t*_F(j)]$$

we have for any $i$, $j$ and $j'$

$$p(i,j)t(i,j) - p(i,j')t(i,j')$$

$$= p*_B(i)p*_F(j)[t*_B(i)+t*_F(j)] - p*_B(i)p*_F(j')[t*_B(i)+t*_F(j')] \qquad (8)$$

Because Equation (7) and Equation (8) have the same left side, the right sides are the same, that is,

$$p_B(i)p_F(j)[t_B(i)+t_F(j)] - p_B(i)p_F(j')[t_B(i)+t_F(j')]$$

$$= p*_B(i)p*_F(j)[t*_B(i)+t*_F(j)] - p*_B(i)p*_F(j')[t*_B(i)+t*_F(j')]$$

Substitute

$$p*_B(i) = cp_B(i)$$

$$p*_F(j) = p_F(j)/c + (c-1)p_D/c.$$

We have

$$p_B(i)p_F(j)[t_B(i)+t_F(j)] - p_B(i)p_F(j')[t_B(i)+t_F(j')]$$

$$= cp_B(i)[p_F(j)/c + (c-1)p_D/c][t*_B(i)+t*_F(j)]$$

$$\quad - cp_B(i)[p_F(j')/c + (c-1)p_D/c][t*_B(i)+t*_F(j')]$$

Equivalently,

$$p_F(j)t_B(i) + p_F(j)t_F(j) - p_F(j')t_B(i) - p_F(j')t_F(j')$$

$$= p_F(j)t*_B(i) + p_F(j)t*_F(j) + (c-1)p_D[t*_F(j)-t*_F(j')]$$

$$\quad - p_F(j')t*_B(i) - p_F(j')t*_F(j')$$

Then we get

$$[t*_B(i)-t_B(i)][p_F(j')-p_F(j)]$$

$$= p_F(j)t*_F(j) + (c-1)p_D[t*_F(j)-t*_F(j')] - p_F(j')t*_F(j')$$

$$\quad - p_F(j)t_F(j) + p_F(j')t_F(j')$$

As a result, for any $j$ such that $p_F(j) \neq p_F(j')$

$t*_B(i) - t_B(i) = \dfrac{p_F(j)t*_F(j) + (c-1)p_D[t*_F(j) - t*_F(j')] - p_F(j')t*_F(j') - p_F(j)t_F(j) + p_F(j')t_F(j')}{p_F(j') - p_F(j)}$. The

left hand side of the above equation cannot change when $j$ changes. So the left hand side must be a constant, denote it as $f$.

Hence, $t*_B(i) = t_B(i) + f$.

Further, the right hand side equals $f$.

$\dfrac{p_F(j)t*_F(j) + (c-1)p_D[t*_F(j) - t*_F(j')] - p_F(j')t*_F(j') - p_F(j)t_F(j) + p_F(j')t_F(j')}{p_F(j') - p_F(j)} = f$.

Hence, there is

$t*_F(j) = \dfrac{p_F(j)t_F(j) + f \cdot [p_F(j') - p_F(j)] + t*_F(j')[p_F(j') + (c-1)p_D] - p_F(j')t_F(j')}{p_F(j) + (c-1)p_D}$

Similarly, according to the assumptions,

$p(i,j)t(i,j) = p_A(i)p_D[t_A(i) + t_D] + p_B(i)p_F(j)[t_B(i) + t_F(j)]$

$p(i,j)t(i,j) = p*_A(i)p*_D[t*_A(i) + t*_D] + p*_B(i)p*_F(j)[t*_B(i) + t*_F(j)]$

for any $i$ such that $p_B(i) \neq 0$ and $p*_B(i) \neq 0$, we have

$$\dfrac{p(i,j)t(i,j)}{p_B(i)} = \dfrac{p_A(i)p_D[t_A(i) + t_D]}{p_B(i)} + p_F(j)[t_B(i) + t_F(j)] \tag{9}$$

$$\dfrac{p(i,j)t(i,j)}{p*_B(i)} = \dfrac{p*_A(i)p*_D[t*_A(i) + t*_D]}{p*_B(i)} + p*_F(j)[t*_B(i) + t*_F(j)] \tag{10}$$

Because $p*_B(i) = cp_B(i)$, we have $\dfrac{p(i,j)t(i,j)}{p*_B(i)} = \dfrac{p(i,j)t(i,j)}{cp_B(i)}$

Also there is $p*_D = p_D$ from Table 2.

So Equation (10) can be transformed into

$$\dfrac{p(i,j)t(i,j)}{p_B(i)} = c \cdot \dfrac{p*_A(i)p_D[t*_A(i) + t*_D]}{p*_B(i)} + c \cdot p*_F(j)[t*_B(i) + t*_F(j)]. \tag{11}$$

Then according to Equation (9), for any $i$ and $i'$ which can be any different value of $i$ with $p_B(i') \neq 0$

$\dfrac{p(i,j)t(i,j)}{p_B(i)} - \dfrac{p(i',j)t(i',j)}{p_B(i')}$

$= \dfrac{p_A(i)p_D[t_A(i) + t_D]}{p_B(i)} + p_F(j)[t_B(i) + t_F(j)] - \dfrac{p_A(i')p_D[t_A(i') + t_D]}{p_B(i')} - p_F(j)[t_B(i') + t_F(j)]$

$= \dfrac{p_A(i)p_D t_A(i)}{p_B(i)} + \dfrac{p_A(i)p_D t_D}{p_B(i)} + p_F(j)[t_B(i) - t_B(i')] - \dfrac{p_A(i')p_D t_A(i')}{p_B(i')} - \dfrac{p_A(i')p_D t_D}{p_B(i')}$

$= \dfrac{p_A(i)p_D t_A(i)}{p_B(i)} - \dfrac{p_A(i')p_D t_A(i')}{p_B(i')} + p_F(j)[t_B(i) - t_B(i')] + \dfrac{p_A(i)p_D t_D}{p_B(i)} - \dfrac{p_A(i')p_D t_D}{p_B(i')}$

$$= \frac{p_A(i)p_D t_A(i)}{p_B(i)} - \frac{p_A(i')p_D t_A(i')}{p_B(i')} + p_F(j)[t_B(i) - t_B(i')] + [\frac{1-p_B(i)}{p_B(i)} - \frac{1-p_B(i')}{p_B(i')}]p_D t_D$$

$$= \frac{p_A(i)p_D t_A(i)}{p_B(i)} - \frac{p_A(i')p_D t_A(i')}{p_B(i')} + p_F(j)[t_B(i) - t_B(i')] + [\frac{1}{p_B(i)} - \frac{1}{p_B(i')}]p_D t_D \qquad (12)$$

Also, according to Equation (11), we have

$$\frac{p(i,j)t(i,j)}{p_B(i)} - \frac{p(i',j)t(i',j)}{p_B(i')}$$

$$= c\frac{p*_A(i)p_D[t*_A(i) + t*_D]}{p*_B(i)} + cp*_F(j)[t*_B(i) + t*_F(j)]$$

$$- c\frac{p*_A(i')p_D[t*_A(i') + t*_D]}{p*_B(i')} - cp*_F(j)[t*_B(i') + t*_F(j)]$$

$$= c\{\frac{p*_A(i)p_D[t*_A(i) + t*_D]}{p*_B(i)} + p*_F(j)[t*_B(i) + t*_F(j)]$$

$$- \frac{p*_A(i')p_D[t*_A(i') + t*_D]}{p*_B(i')} - p*_F(j)[t*_B(i') + t*_F(j)]\}$$

$$= c\{\frac{p*_A(i)p_D t*_A(i)}{p*_B(i)} + \frac{p*_A(i)p_D t*_D}{p*_B(i)} + p*_F(j)[t*_B(i) - t*_B(i')]$$

$$- \frac{p*_A(i')p_D t*_A(i')}{p*_B(i')} - \frac{p*_A(i')p_D t*_D}{p*_B(i')}\}$$

$$= c\{\frac{p*_A(i)p_D t*_A(i)}{p*_B(i)} - \frac{p*_A(i')p_D t*_A(i')}{p*_B(i')} + p*_F(j)[t*_B(i) - t*_B(i')]$$

$$+ \frac{p*_A(i)p_D t*_D}{p*_B(i)} - \frac{p*_A(i')p_D t*_D}{p*_B(i')}\}$$

$$= c\{\frac{p*_A(i)p_D t*_A(i)}{p*_B(i)} - \frac{p*_A(i')p_D t*_A(i')}{p*_B(i')} + p*_F(j)[t*_B(i) - t*_B(i')]$$

$$+ [\frac{1-p*_B(i)}{p*_B(i)} - \frac{1-p*_B(i')}{p*_B(i')}]p_D t*_D\}$$

$$= c\{\frac{p*_A(i)p_D t*_A(i)}{p*_B(i)} - \frac{p*_A(i')p_D t*_A(i')}{p*_B(i')} + p*_F(j)[t*_B(i) - t*_B(i')]$$

$$+ [\frac{1}{p*_B(i)} - \frac{1}{p*_B(i')}]p_D t*_D\}$$

$$(13)$$

Substitute the following values into Equation (13),

$$p*_B(i) = cp_B(i),$$

$$p*_F(j) = p_F(j)/c + (c-1)p_D/c,$$

$$t*_B(i) = t_B(i) + f.$$

Then

$$\frac{p(i,j)t(i,j)}{p_B(i)} - \frac{p(i',j)t(i',j)}{p_B(i')}$$

$$= c\{\frac{p*_A(i)p_D t*_A(i)}{cp_B(i)} - \frac{p*_A(i')p_D t*_A(i')}{cp_B(i')} + [p_F(j)/c + (c-1)p_D/c][t_B(i) - t_B(i')]$$

$$+ [\frac{1}{cp_B(i)} - \frac{1}{cp_B(i')}]p_D t*_D\}$$

$$= \frac{p*_A(i)p_D t*_A(i)}{p_B(i)} - \frac{p*_A(i')p_D t*_A(i')}{p_B(i')} + [p_F(j) + (c-1)p_D][t_B(i) - t_B(i')]$$

$$+ [\frac{1}{p_B(i)} - \frac{1}{p_B(i')}]p_D t*_D \qquad (14)$$

Because Equation (12) and Equation (14) have the same left side, the right sides are the same, that is,

$$\frac{p_A(i)p_D t_A(i)}{p_B(i)} - \frac{p_A(i')p_D t_A(i')}{p_B(i')} + p_F(j)[t_B(i) - t_B(i')] + [\frac{1}{p_B(i)} - \frac{1}{p_B(i')}]p_D t_D$$

$$= \frac{p*_A(i)p_D t*_A(i)}{p_B(i)} - \frac{p*_A(i')p_D t*_A(i')}{p_B(i')} + [p_F(j) + (c-1)p_D][t_B(i) - t_B(i')]$$

$$+ [\frac{1}{p_B(i)} - \frac{1}{p_B(i')}]p_D t*_D$$

Equivalently,

$$\frac{p_A(i)p_D t_A(i)}{p_B(i)} - \frac{p_A(i')p_D t_A(i')}{p_B(i')} + p_F(j)[t_B(i) - t_B(i')] - \frac{p*_A(i)p_D t*_A(i)}{p_B(i)}$$

$$+ \frac{p*_A(i')p_D t*_A(i')}{p_B(i')} - [p_F(j) + (c-1)p_D][t_B(i) - t_B(i')]$$

$$= [\frac{1}{p_B(i)} - \frac{1}{p_B(i')}]p_D t*_D - [\frac{1}{p_B(i)} - \frac{1}{p_B(i')}]p_D t_D$$

The left side can be simplified into

$$\frac{p_D}{p_B(i)}\{p_A(i)t_A(i) - p*_A(i)t*_A(i)\}$$

$$- \frac{p_D}{p_B(i')}\{p_A(i')t_A(i') - p*_A(i')t*_A(i')\} + [(1-c)p_D][t_B(i) - t_B(i')]$$

Meanwhile, the right side can be simplified into

$$[\frac{1}{p_B(i)} - \frac{1}{p_B(i')}]p_D(t*_D - t_D)$$

Then we have,

$$\frac{p_D}{p_B(i)}\{p_A(i)t_A(i) - p*_A(i)t*_A(i)\} - \frac{p_D}{p_B(i')}\{p_A(i')t_A(i') - p*_A(i')t*_A(i')\}$$

$$+[(1-c)p_D][t_B(i)-t_B(i')]$$

$$=[\frac{1}{p_B(i)}-\frac{1}{p_B(i')}]p_D(t*_D-t_D)$$

So for any $i$ such that $p_B(i)\neq p_B(i')$, there is,

$t*_D-t_D=$

$$\frac{\frac{p_D}{p_B(i)}\{p_A(i)t_A(i)-p*_A(i)t*_A(i)\}-\frac{p_D}{p_B(i')}\{p_A(i')t_A(i')-p*_A(i')t*_A(i')\}+[(1-c)p_D][t_B(i)-t_B(i')]}{[\frac{1}{p_B(i)}-\frac{1}{p_B(i')}]p_D}$$ The

left hand side of the above equation cannot change when $i$ changes. So the left hand side must be a constant, denote it as $e$.

Hence, $t*_D=t_D+e$

Because

$$p(i,j)t(i,j)=p_A(i)p_D[t_A(i)+t_D]+p_B(i)p_F(j)[t_B(i)+t_F(j)]$$

$$p(i,j)t(i,j)=p*_A(i)p*_D[t*_A(i)+t*_D]+p*_B(i)p*_F(j)[t*_B(i)+t*_F(j)]$$

There is

$$p_A(i)p_D[t_A(i)+t_D]+p_B(i)p_F(j)[t_B(i)+t_F(j)]$$

$$=p*_A(i)p*_D[t*_A(i)+t*_D]+p*_B(i)p*_F(j)[t*_B(i)+t*_F(j)]$$

Substitute

$p*_B(i)=cp_B(i)$,

$p*_F(j)=p_F(j)/c+(c-1)p_D/c$,

$t*_B(i)=t_B(i)+f$,

$$t*_F(j)=\frac{p_F(j)t_F(j)+f\cdot[p_F(j')-p_F(j)]+t*_F(j')[p_F(j')+(c-1)p_D]-p_F(j')t_F(j')}{p_F(j)+(c-1)p_D},$$

and

$t*_D=t_D+e$,

We get

$$p_A(i)p_D[t_A(i)+t_D]+p_B(i)p_F(j)[t_B(i)+t_F(j)]$$

$$=[1-cp_B(i)]p_D[t*_A(i)+t_D+e]+cp_B(i)[p_F(j)/c+(c-1)p_D/c]\times$$

$$\{t_B(i)+f+\frac{p_F(j)t_F(j)+f\cdot[p_F(j')-p_F(j)]+t*_F(j')[p_F(j')+(c-1)p_D]-p_F(j')t_F(j')}{p_F(j)+(c-1)p_D}\}$$

Equivalently,

$$[1-cp_B(i)]p_D[t*_A(i)+t_D+e]-[1-p_B(i)]p_D[t_A(i)+t_D]$$

$$=p_B(i)p_F(j)[t_B(i)+t_F(j)]-cp_B(i)[p_F(j)/c+(c-1)p_D/c]\times\{t_B(i)+f$$

$$+\frac{p_F(j)t_F(j)+f[p_F(j')-p_F(j)]+t*_F(j')[p_F(j')+(c-1)p_D]-p_F(j')t_F(j')}{p_F(j)+(c-1)p_D}\}$$

The left side can be simplified into

$$p_D\{[1-cp_B(i)]t*_A(i)-p_A(i)t_A(i)+(1-c)p_B(i)t_D-cep_B(i)+e\}$$

The right side can be simplified into

$$p_B(i)\{t_B(i)(1-c)p_D + [f + t*_F(j')][(1-c)p_D - p_F(j')] + p_F(j')t_F(j')\}$$

Hence,

$$p_D\{[1-cp_B(i)]t*_A(i) - p_A(i)t_A(i) + (1-c)p_B(i)t_D - cep_B(i) + e\}$$

$$= p_B(i)\{t_B(i)(1-c)p_D + [f + t*_F(j')][(1-c)p_D - p_F(j')] + p_F(j')t_F(j')\}$$

Then we have,

$$t*_A(i)[1-cp_B(i)]$$

$$= p_B(i)\{t_B(i)(1-c) + [f + t*_F(j')][1-c-\frac{p_F(j')}{p_D}] + \frac{p_F(j')t_F(j')}{p_D}\}$$

$$+ p_A(i)t_A(i) - (1-c)p_B(i)t_D + cep_B(i) - e$$

$$= p_A(i)t_A(i) + p_B(i)\{[t_B(i)-t_D](1-c) + [f + t*_F(j')]\times$$

$$[1-c-\frac{p_F(j')}{p_D}] + \frac{p_F(j')t_F(j')}{p_D} + ce\} - e$$

So we get,

$$t*_A(i) = \frac{p_A(i)t_A(i) + p_B(i)\{[t_B(i)-t_D](1-c) + [f + t*_F(j')][1-c-\frac{p_F(j')}{p_D}] + \frac{p_F(j')t_F(j')}{p_D} + ce\} - e}{1-cp_B(i)}$$

II. Conversely, suppose for all $1\le i \le I$, and $1\le j \le J$, there exist $p*_A, p*_B, p*_D, p*_F(j), t*_A, t*_B, t*_D(i), t*_F(j)$ such that

$$0 \le p*_A, p*_B, p*_D, p*_F(j) \le 1$$

$$t*_A, t*_B, t*_D(i), t*_F(j) \ge 0$$

with the following equations,

$$p*_B(i) = cp_B(i),$$

$$p*_F(j) = p_F(j)/c + (c-1)p_D/c,$$

$$p*_D = p_D$$

$$t*_B(i) = t_B(i) + f,$$

$$t*_F(j) = \frac{p_F(j)t_F(j) + f\cdot[p_F(j')-p_F(j)] + t*_F(j')[p_F(j')+(c-1)p_D] - p_F(j')t_F(j')}{p_F(j)+(c-1)p_D},$$

$$t*_D = t_D + e,$$

and

$$t*_A(i) = \frac{p_A(i)t_A(i) + p_B(i)\{[t_B(i)-t_D](1-c) + [f + t*_F(j')][1-c-\frac{p_F(j')}{p_D}] + \frac{p_F(j')t_F(j')}{p_D} + ce\} - e}{1-cp_B(i)}$$

as well as the bounds for $c$ in Table 3.

Then

$$p*_A p*_D [t*_A(i) + t*_D]$$

$$= [1 - p*_B(i)]p*_D [t*_A(i) + t*_D]$$

$$= \ p_A(i)t_A(i)p_D + p_B(i)\{[t_B(i)-t_D](1-c)p_D + [f + t*_F(j')][(1-c)p_D - p_F(j')]$$
$$+ \ p_F(j')t_F(j') + cep_D\} - ep_D + (t_D + e)[1 - cp_B(i)]p_D$$

$$= \ p_A(i)p_D[t_D + t_A(i)] + p_B(i)\{t_B(i)(1-c)p_D + [f + t*_F(j')][(1-c)p_D - p_F(j')] + p_F(j')t_F(j')\}$$

Meanwhile,

$$p*_B(i)p*_F(j)[t*_B(i) + t*_F(j)]$$

$$= \ cp_B(i)[p_F(j)/c + (c-1)p_D/c] \times [t_B(i) + f$$
$$+ \frac{p_F(j)t_F(j) + f \cdot [p_F(j') - p_F(j)] + t*_F(j')[p_F(j') + (c-1)p_D] - p_F(j')t_F(j')}{p_F(j) + (c-1)p_D}]$$

$$= \ p_B(i)\{[t_B(i) + f][p_F(j) + (c-1)p_D] + p_F(j)t_F(j) + f \cdot [p_F(j') - p_F(j)]$$
$$+ t*_F(j')[p_F(j') + (c-1)p_D] - p_F(j')t_F(j')\}$$

$$= \ p_B(i)\{t_B(i)[p_F(j) + (c-1)p_D] + p_F(j)t_F(j) + [t*_F(j') + f][p_F(j') + (c-1)p_D]$$
$$- p_F(j')t_F(j')\}$$

Putting these two parts together, we get

$$p*_A \, p*_D [t*_A(i) + t*_D] + p*_B(i)p*_F(j)[t*_B(i) + t*_F(j)]$$

$$= \ p_A(i)p_D[t_D + t_A(i)] + p_B(i)\{t_B(i)(1-c)p_D + [f + t*_F(j')][(1-c)p_D - p_F(j')]$$
$$+ p_F(j')t_F(j')\}$$

$$+ \ p_B(i)\{t_B(i)[p_F(j) + (c-1)p_D] + p_F(j)t_F(j) + [t*_F(j') + f][p_F(j') + (c-1)p_D]$$
$$- p_F(j')t_F(j')\}$$

$$= \ p_A(i)p_D[t_A(i) + t_D] + p_B(i)p_F(j)[t_B(i) + t_F(j)]$$

$$= \ p(i,j)t(i,j)$$

□

Remarks on nonnegative measure values.

     In some applications the measure associated with an arc may be positive or negative. For example, the measure in a decision tree is a payoff, which could be positive (a gain) or negative (a loss). For application to response time, we assume the measure associated with an arc is a time, a nonnegative quantity. In such an application, an admissible transformation of a measure must transform a nonnegative quantity to another nonnegative quantity.

     The bounds in Table 4 on measure scaling parameters achieve this.

Table 4

*Bounds on Scaling Parameters in Admissible Transformations*

*For the Standard Binary Tree for Ordered Processes*

*For Measure Parameters to be Nonnegative*

_______________________________________

$$\max\{-t_B(i)\} \leq f \leq \min\{t^*_B(i)\}$$

$$\max\{-t_D\} \leq e \leq \min\{t^*_D\}$$

$$\max\{-t_C\} \leq k \leq \min\{t^*_C\}$$

_______________________________________

From Table 2, consider the admissible transformation $t^*_B(i) = t_B(i) + f$. For both $t^*_B(i)$ and $t_B(i)$ to be nonnegative, $f$ has boundaries as

$$t^*_B(i) = t_B(i) + f \geq 0$$
$$t_B(i) = t^*_B(i) - f \geq 0$$

So for $f$, there is

$$\max\{-t_B(i)\} \leq f \leq \min\{t^*_B(i)\}.$$

From Table 2, consider the admissible transformation $t^*_D = t_D + e$. For both $t^*_D$ and $t_D$ to be nonnegative, $e$ has boundaries as

$$t^*_D = t_D + e \geq 0$$
$$t_D = t^*_D - e \geq 0$$

So for $e$, there is

$$\max\{-t_D\} \leq e \leq \min\{t^*_D\}.$$

*Degrees of freedom.* Using the admissible transformations of parameters allows us to calculate the degrees of freedom for the Standard Binary Tree for Ordered processes in the following corollary.

**Corollary 3.** *Suppose probability matrix* $\mathbf{P} = (p(i,j))$, *correct-response-measure matrix* $\mathbf{T} = (t(i,j))$ *and incorrect-response-measure-matrix* $\mathbf{T_w} = (t_w(i,j))$ *are produced by Factors* $\Phi$ *and* $\Psi$ *selectively influencing two vertices ordered by the factors in the Standard Binary Tree for*

*Ordered Processes, with the vertex selectively influenced by Factor Φ preceding the vertex selectively influenced by Factor Ψ, with probability parameters $p_A(i)$, $p_B(i)$, $p_C$, $p_D$, $p_E(j)$, and $p_F(j)$, and measure parameters $t_A(i)$, $t_B(i)$, $t_C$, $t_D$, $t_E(j)$, and $t_F(j)$.*

*Suppose Factor Φ has I levels and Factor Ψ has J levels. Then the degrees of freedom are $3IJ - 3I - 3J + 3$.*

**Proof.** For each combination of a level $i$ of Factor Φ and a level $j$ of Factor Ψ we have an observed probability of a correct response, an observed measure for a correct response and an observed measure for an incorrect response. The probability of an incorrect response is determined by the probability of a correct response. Hence the total number of observations is $3IJ$. The arc probabilities to be estimated are $p_A(i)$, $p_B(i)$, $p_C$, $p_D$, $p_E(j)$, and $p_F(j)$. But for every $i$

$$p_A(i) + p_B(i) = 1$$

and for every $j$,

$$p_E(j) + p_F(j) = 1.$$

Hence the number of independent arc probabilities to be estimated is $I + J + 1$. The arc measure parameters to be estimated are $t_A(i)$, $t_B(i)$, $t_C$, $t_D$, $t_E(j)$, and $t_F(j)$; their number is $2I + 2 + 2J$. The scaling parameters to be freely selected are $c$ for arc probabilities, $e$, $f$, and $k$ for measures, also $t^*_E(j')$ and $t^*_F(j')$ so there are 6 scaling parameters.

Hence, there are $3IJ-(I+J+1)-(2+2I+2J)+6 = 3(IJ - I - J + 1)$ degrees of freedom.

□

## Discussion

Multinomial Processing Trees are widely used as models of phenomena in psychology (reviewed by Batchelder & Riefer, 1999; Erdfelder, Auer, Hilbig, Aβfalg, Moshagen & Nadarevic, 2009; and Hütter & Klauer, 2016). One reason is straightforwardness and relative simplicity, but the major reason is their ability to often account for data. Agreement with data is usually evaluated by goodness of fit. Additional support is sometimes provided by a factorial experiment, with tests of whether factors selectively influence vertices in an MPT (reviewed by Schweickert, Fisher & Sung, 2012).

In a binary MPT exactly two arcs descend from each nonterminal vertex. MPTs with more than two arcs descending from a vertex are occasionally inferred from data, for a recent example, see Schweickert, Dhir, Zheng and Poirier (2020). But many MPTs currently in use are binary; most in the applications discussed in the reviews cited above are binary trees.

Here we provide necessary and sufficient conditions, which can be tested with data from factorial experiments, for selective influence of the factors on vertices in a a particular binary MPT, the Standard Binary Tree for Ordered Processes. This MPT has a special role, because under certain conditions if each of two factors selectively influences a different vertex in an arbitrary MPT, that MPT is equivalent for the factors to this one. Methods of testing are beyond the scope of this paper, but described in Schweickert and Zheng (2018). Parameter values are not unique. Admissible transformations are given that allow one set of parameter values to be transformed to another. Degrees of freedom for an experiment with the two factors are calculated from the number of observations and the number of parameters to be estimated, both depend on the number of levels of the factors.

*Appendix*

Theorem 5 of Schweickert and Zheng (2019b)

*Suppose for all i and j, $0 < p(i, j) < 1$. Probability matrix $\mathbf{P} = (p(i, j))$, correct-response measure matrix $\mathbf{T} = (t(i, j))$, and incorrect-response-measure matrix $\mathbf{Tw} = (tw(i, j))$ are produced by Factor $\Phi$ and Factor $\Psi$ selectively influencing two different vertices ordered by the factors in the Standard Tree for Ordered Processes, with the vertex selectively influenced by Factor $\Phi$ preceding the vertex selectively influenced by Factor $\Psi$, if and only if there is a level n of Factor $\Psi$ and for every level i of Factor $\Phi$ there are numbers $r_i \geq 0$ and $s_i$ such that the following three conditions are true.*

*1. The columns of P can be numbered so $j \geq j'$ implies that for every i, $p(i, j) \geq p(i, j')$.*

*2. There are levels $i^*$ and $j^*$ such that for every i and j,*
$$p(i, j) - p(i, j^*) = r_i[p(i^*, j) - p(i^*, j^*)].$$

*3. Let $\max\{r_i\} = r_h$. For every j,*
$$r_h r_i s_i[p(h, j) - p(h, n)]$$
$$= r_h[p(i, j)t(i, j) - p(i, n)t(i, n)]$$
$$\quad - r_i[p(h, j)t(h, j) - p(h, n)t(h, n)]$$
$$= -r_h\{[1 - p(i, j)]t_w(i, j) - [1 - p(i, n)]t_w(i, n)\}$$
$$\quad + r_i\{[1 - p(h, j)]t_w(h, j) - [1 - p(h, n)]t_w(h, n)\}.$$